# Formal Process Virtual Machine for Smart Contracts Verification

Zheng Yang[a,*], Hang Lei[a]

[a]*School of Information and Software Engineering, University of Electronic Science and Technology of China,
No.4, Section 2, North Jianshe Road, 610054, Sichuan, Chengdu, P.R. China.*

**Abstract**

This paper reports on the development and verification of a novel formal symbolic process virtual machine (FSPVM) for verifying the reliability and security of Ethereum smart contracts, denoted as FSPVM-E, in Coq proof assistant. It adopts execution-verification isomorphism (EVI), an extension of Curry-Howard isomorphism (CHI), as its fundamental theoretical framework. The current version of FSPVM-E is constructed on a general, extensible, and reusable formal memory (GERM) framework, an extensible and universal formal intermediate programming language Lolisa, which is a large subset of the Solidity programming language using generalized algebraic datatypes, and the corresponding formally verified interpreter of Lolisa, denoted as FEther. It supports the ERC20 standard and can automatically simultaneously symbolically execute the smart contract programs of Ethereum and verify their reliability and security properties using Hoare logic in Coq. In addition, this work, contributes to solving the problems of automation, inconsistency and reusability in higher-order logic theorem proving.

## 1. Introduction

In recent years, with the development of computer science, a number of software programs have been deployed in many critical domains. The smart contract program of blockchain technology [1] is one notable example of critical software. A smart contract is a kind of digital contract where the code is the law. However, this feature of smart contracts makes them susceptible to attack that can result in economic loss. Therefore, verifying the reliability and security of such programs in the most rigorous manner available is crucial. Higher-order logic theorem proving (HOLTP) is one of the most rigorous technologies for verifying the reliability and security of programs to build trustworthy software systems. However, in applying HOLTP, it is necessary to abstract a specific formal model for the target software system manually, with the help of proof assistants [3]. Although this type of formal verification technology has many advantages, such as providing sufficient flexibility in designing formal models, abstracting and expressing very complex systems, the formal models are dependent on the experience, knowledge, and proficiency of the users. Thus, the standard approaches have a low-level of automation, are prone to inconsistencies and have a low-level of reusability, which presents a barrier to the extensive application of HOLTP.

One possible solution for addressing these problems is to design an extensible and universal formal symbolic process virtual machine (FSPVM) like KLEE [6], but developed in a HOLTP system, which can symbolically execute real world programs and verify their properties automatically using the results of the execution.

This paper gives a high-level overview of our current research, highlighting two contributions. First, we present a novel extension of Curry-Howard isomorphism (CHI) [7], denoted as execution-verification isomorphism (EVI) [4] which integrates the advantages of model checking and theorem proving technology. EVI is the basic theory for combining higher-order logic systems (HOLS), supporting CHI, and symbolic execution technology to extend and virtualize HOLS as an FSPVM to solve the problems of automation, consistency and reusability in higher-order logic theorem proving. Second, we take EVI as the fundamental theoretical framework to develop an FSPVM for Ethereum (FSPVM-E) including a general, extensible and reusable formal memory framework called GERM, an extensible and universal formal intermediate programming language, denoted as Lolisa (the present version is a large subset of the Solidity programming language), which automates generalized algebraic datatypes, and the corresponding formally verified interpreter of Lolisa, denoted as FEther. The FSPVM-E is entirely developed in Coq [8], which is one of the best higher-order logic theorem proving assistants, based on the Calculus of inductive construction (Cic) that supports CHI. The FSPVM-E is employed to automatically complete the formal verification of the security and reliability properties of Ethereum smart contract programs. To our knowledge, our work is the first to automatically validate the formal syntax and semantics of the Solidity programming language, and systematically build a virtual execution and verification environment entirely in Coq for automatically verifying Ethereum smart contracts, solving the problem of inconsistency.

* Corresponding author: Zheng Yang.
  E-mail address: zyang.uestc@gmail.com

The remainder of this paper is organized as follows. Section 2 introduces related work on consistency, reusability and automation as well as formal verification for Ethereum smart contracts. Section 3 briefly illustrates the basic concepts and advantages of EVI. Section 4 describes the overall implementation of FSPVM-E and simple cases of its application. Finally, Section 5 presents preliminary conclusions and directions for future work.

**2. Related work**

Program verification using higher-order logic theorem proving is a very important theoretical field in computer science. Many researchers have addressed the problems of consistency, reusability, and automation from various aspects and have developed new tools to contribute to this field. For the problems of consistency and reusability, one well-known and efficient method is to formalize real world programming languages in an intermediate programming language (IPL) and design a formal memory model as the state model. Since the late 1960's, many studies have focused on building memory models mathematically for program verification. One of the milestones was the CompCert project on compiler verification in 2008 [9]. The team of CompCert formalized an equivalent IPL called Clight for the C programming language using Coq. They also developed a formal memory model for low-level imperative languages such as C and compiler intermediate languages. These works have served as the basis for some interesting and powerful program verification and analysis frameworks. Verified software toolchain (VST) [10] and deep specifications [11] are two representative projects that have been developed in conjunction with the IPL and formal memory model provided by CompCert. However, these works have focused on specific domains and programming languages, and their formal memory models are deeply embedded in their framework, making it difficult to extend and modify them for supporting different high-level specifications, which would enable the formalization of programs written in different high-level languages. Also, most of them are focused on a subset of problems in consistency, reusability, or automation instead of considering them all simultaneously. Furthermore, it requires expert knowledge to rebuild the source code of the programs and construct the abstract layers manually. Hence, there is still a risk of inconsistency and it is not possible to fully utilize automated theorem proving technologies.

Meanwhile, formal verification for blockchain technology has become a subject of particular interest in recent years and many prominent research works have focused on the formal verification of the bytecode of the Ethereum virtual machine (EVM). For example, KEVM [12] is a formal semantics for the EVM written using the K-framework, like the formalization conducted in Lem [13]. However, the development of a high-level formal specification for Solidity and relevant formal verification tools have attracted considerably less interest from researchers despite its importance for programming and debugging smart contract software. Our present study fills this gap.

**3. EVI**

The concept of EVI is applied here to increase the degree to which the process of program verification is conducted automatically by combining HOLTP and symbolic execution. EVI is the abstract fundamental theoretical framework for extending and virtualizing a higher-order logic formal system, supporting CHI to become a universal and extensible formal symbolic process virtual machine, which addresses the consistency, reusability and automation problems of HOLTP simultaneously. The basic concept of EVI and its main features will be briefly introduced in following subsections, while further details can be found in our other work [4].

*3.1. Conceptual basis of EVI*

EVI has two core elements. The first is the theoretical basis of extending a formal logic system which supports CHI theory as a virtual formal symbolic execution environment. The second is the isomorphism between symbolic execution and formal verification of HOLTP.

Proof assistants are a kind of specialized software, which provide an environment for developing mathematical constructs. In this environment, programmers can use the vernacular to define mathematical objects and finally write proofs. They are evaluated by the trusted core of the proof assistant which generates the logic results. Hence, compared with the architecture of a PVM, they provide a fundamental virtual environment, and we can extend HOLTP assistants, like Coq, as a special FSPVM, which contains the minimal full virtualization environment that can symbolically execute programs written in a high-level general-purpose programming language $\mathcal{L}$.

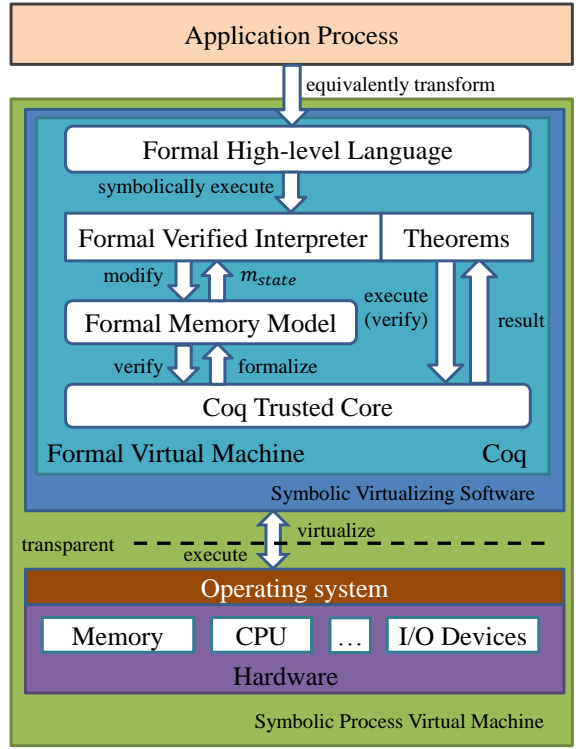

Figure 1. Overall architecture of FSPVM in Coq

Specifically, we take Coq as an example because Coq is one of the best proof assistants whose basic theory Cic supports CHI. As Rule 1, the logic environment $\mathcal{E}$ of Coq can be viewed as the abstraction or virtualization of the real world physical operating environment, which provides a functional programming language (FPL) called Gallina.

$$physical\ operating\ environment \xrightarrow{virtualizes} formal\ logic\ environment \qquad (1)$$

This virtual environment includes two sub-environments: a functional-programming environment and a proof editing environment [14], which are used to define formal functions and proof obligations. In Rules 2 and 3, Gallina plays the role of low-level machine code in this $\mathcal{E}$ and the trusted core of Coq (TCOC) can be seen as the virtualization of the CPU, which can evaluate (be isomorphic to proving) the programs, also called proof terms, written in Gallina.

$$CPU \xrightarrow{virtualizes} TCOC \qquad (2)$$
$$native\ machine\ code \xrightarrow{virtualizes} Gallina \qquad (3)$$

However, this fundamental environment provided by Coq is not sufficient to symbolically execute $\mathcal{L}$, because the TCOC only has two functions, evaluating and proving, so it cannot parse and execute a real-world program (RWprogram) written in $\mathcal{L}$. Therefore, it must be extended. As in the blueprint given in our paper [4], if we want to extend Coq to become an FSPVM that can symbolically execute and verify real world programs, in addition to the fundamental logic environment of Gallina and TCOC, we need to construct three more key elements: a formal general memory model (FMemory), a high-level formal intermediate language $\mathcal{FL}$, which is equivalent to $\mathcal{L}$, for rewriting the formal version of RWprogram as FRWprogram, and the respective formally verified interpreter (FInterpreter). As illustrated in Rules 4 and 5, the FMemory simulates the real-world physical memory space and operations, while the $\mathcal{FL}$ is the equivalent formal version of $\mathcal{L}$, which can be analyzed in $\mathcal{E}$ directly.

$$physical\ memory \xrightarrow{virtualizes} formal\ memory\ model \quad (4)$$

$$\left(\mathcal{L} \xrightarrow{formalizes} \mathcal{FL}\right) \wedge (\mathcal{L} \equiv \mathcal{FL}) \supset RWprogram \equiv FRWprogram \quad (5)$$

The real-world program RWprogram written in $\mathcal{L}$ can be executed directly with the help of corresponding interpreters in a physical operating environment. Compared with the physical operating environment, the combination of FMemory and $\mathcal{E}$ has already been virtualized in a minimal higher-order logic operating environment. Therefore, although a FRWprogram rewritten in the $\mathcal{FL}$ cannot be executed directly in an $\mathcal{E}$, such as Coq, we can implement an FInterpreter using FPL based on the higher-order logic operating environment that follows the formal syntax and semantics of the corresponding $\mathcal{FL}$ to simulate the execution process of the RWprogram in the real world and interpret the FRWprogram so that it can be symbolically executed in $\mathcal{E}$ directly with the same process as is conducted in the real world. This process is illustrated in Figure 2.

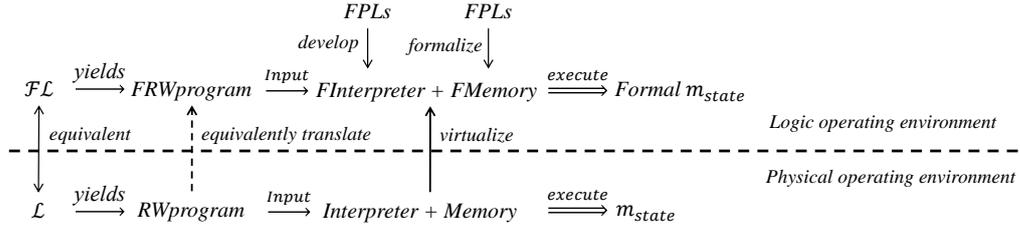

Figure 2. Equivalence between real world program (RWprogram) execution and execution in a logic environment

Then, because the FInterpreter is a kind of *program* written in an FPL, such as Gallina provided by Coq, which supports CHI, the evaluation process of FInterpreter with abstract definition as

$$formal\ interpreter: memory \rightarrow FRWprogram \rightarrow option\ memory$$

satisfies CHI Rules 6 and 7 [15].

$$propositions\ as\ types \quad (6)$$
$$proofs\ as\ evaluation\ of\ programs \quad (7)$$

Meanwhile, as illustrated in Figure 2, when the FInterpreter takes the FRWprogram and the state of FMemory as arguments, the evaluation process of the FInterpreter is equivalent to the symbolic execution of the FRWprogram. Hence, we have Rule 8 that the *proofs* are isomorphic with the *execution of programs*. Then, as Rule 9 and 10, in Coq or similar proof assistants, propositions strengthened by special limitations, such as Hoare logic, are the properties for special conditions, and the respective proofs are the verifications of properties.

$$proofs\ correspond\ to\ evaluation\ of\ programs\ correspond\ to\ execution\ of\ programs \quad (8)$$
$$properties\ correspond\ to\ stronger\ propositions\ correspond\ to\ type \quad (9)$$
$$proofs\ correspond\ to\ verification \quad (10)$$

Therefore, we have Rules 11 and 12 which show that program verification in higher-order theorem-proving assistants, which support CHI, is isomorphic with symbolic execution in the logic operating environment built using the proof assistants.

$$properties\ correspond\ to\ stronger\ propositions \quad (11)$$
$$execution\ corresponds\ to\ verification \quad (12)$$

Here, the EVI combines the formal verification of higher-order theorem proving and symbolic execution technology.

*3.2. Main features*

The proposed FSPVM based on EVI is able to solve the consistency, reusability, and automation problems that are present in the standard formal verification approaches of HOLTP.

Addressing consistency problems, we note that, according to Rule 12, the execution of FRWprograms written in an $\mathcal{FL}$ is isomorphic to their formal verification. Therefore, FRWprograms obviously represent the formal models of the corresponding RWprograms. According to Rule 5, RWprogram ≡ FRWprogram . This means the formal model is equivalent to the target RWprogram, without any consistency problems. In addition, since $\mathcal{L} \equiv \mathcal{FL}$, the modeling process of formalization is standardized as equivalently translating RWprograms into FRWprograms line by line mechanically without the need for rebuilding, abstracting, or any other steps that would depend on the experience, knowledge, and proficiency of users. Therefore, no consistency problems will be introduced during the construction of the formal models. Furthermore, all the elements of the FSPVM are defined in the higher-order logic proof assistants directly, using the FPLs provided by the proof assistants. Thus, properties such as reliability, security, and functional correctness, can be certified in the proof assistants without any additional process of modelling or abstracting.

For reusability problems, because of the symbolic execution in proof assistants, the execution results are logic expressions. Therefore these results can be directly employed to be verified in any theorems. Moreover, the sets of results, which have been verified, can be directly applied in other theorem verifications. Besides, as mentioned above, FRWprograms are the formal models of corresponding RWprograms, and FRWprograms can be executed for verifying any relevant theorems rather than rebuilding the entire formal models.

Automation problems should be considered from two perspectives. In modeling, because FRWprograms are the formal models of corresponding RWprograms, the formalization process of building formal models is identical with the process of rewriting the RWprograms $\mathcal{FL}$ using mechanically. And this mechanical translation process can be conducted by specific translators automatically, and thereby reduce the workload associated with the building of formal models. In verification, as mentioned above, formal verification can be conducted automatically by symbolically executing FRWprograms in FSPVM. And, obviously, the program verification process of all formal models based on EVI has been unified as the process of evaluating FInterpreter $(m_{state}, FRWprogram)$, and proving the equivalence between the result memory state and the excepted final memory state. Thus, the differences between the program verification processes among different formal models have been reduced. It then becomes possible to design subtactics based on the tactic mechanism provided by proof assistants that can conduct different parts of the verification process, and combine them in a larger tactic. In this manner, the verification process can be conducted in a fully automatic fashion by employing a combination of tactics.

## 4. Overview of FSPVM-E

Taking EVI as the basic theory, a general, extensible, and reusable FSPVM for the formal verification of Ethereum smart contracts are built and verified entirely in Coq, denoted as FSPVM-E. And FSPVM-E has already been able to semi-automatically symbolically execute Ethereum smart contracts written in Solidity and verify the properties about the reliability, security and functional correctness simultaneously.

*4.1. Architecture*

The FSPVM-E framework is implemented entirely in Coq, and the FSPVM of EVI is the fundamental theoretical framework for building FSPVM-E. The overall structure of FSPVM-E is strictly following Figure 1 that is constructed by three parts. To be specific, the basic element of FSPVM is a general, extensible, and reusable formal memory (GERM) framework that can simultaneously support different formal verification specifications, particularly at the code level. The framework simulates physical memory hardware structure, including a low-level formal memory space, and provides a set of simple, nonintrusive application programming interfaces and assistant tools using Coq that can support different formal verification specifications simultaneously. The proposed GERM framework is independent and customizable, and was verified entirely in Coq. The second part is a formal intermediate programming language, denoted as Lolisa, which is a large subset of the Solidity programming language mechanized in Coq. And the semantics of Lolisa are based on GERM framework. The third part is a formal verified interpreter for Lolisa, called FEther, which connect GERM, Lolisa and TCOC together to symbolically execute and verify the smart contracts of Ethereum. These three parts are introduced specifically in Subsections 4.2–4.4.

*4.2. General formal memory model*

In [4], we have developed a general, extensible, and reusable formal memory framework based on higher-order logic using Coq, denoted as GERM, and it is served as the basis of FSPVM. To be specific, The GERM framework is designed and implemented based on Cic, which is well suited as a basis for arbitrary high-level specifications in different formal models for program verification. Specifically, the GERM framework can be reused with different program verification formal models to store and generate intermediate states.

The overall GERM framework structure is illustrated in Fig. 1. According to the figure, the GERM framework comprises two main components: a formal memory model in a trusted domain and assistant tools in a general domain. The formal memory model includes three levels from bottom to top: a formal memory space, low-level memory management operations, and basic memory management APIs. Assistant tools are employed in the GERM framework to obtain user requirements and generate dynamic specifications. Particularly, although these assistant tools are implemented in the general domain using general-purpose programming languages, the relation between the assistant tools and the respective results satisfies the non-aftereffect property, discussed in [4]. As such, the verified results are not influenced by the assistant tools implementation.

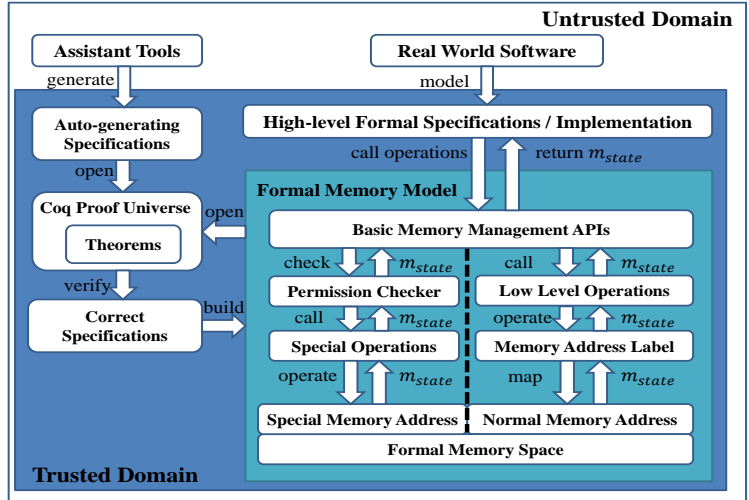

Figure 3. Architecture of the GERM framework

The workflow of the GERM framework can be defined in conjunction with Figure 3 as follows. A user first sets initial requirements, such as memory size, and then the assistant tools generate the respective specifications. Next, the entire formal memory model is certified according to the correctness properties employed in Coq. In the Coq specification, the judgments of the dynamic semantics are encoded as mutually inductive predicates, and the functions are written in Gallina, which is a non-Turing complete language that has eliminated halting problems. If the formal memory model satisfies all required properties, then the specific GERM framework has been constructed successfully. The user can then build a high-level formal model based on the generated GERM framework. The complete workload for constructing the GERM framework with 100 memory blocks is about 3000 Coq lines and 100 C++ lines.

*4.3. Formal intermediate programming language*

The source language of the current version FSPVM, called Lolisa [5], is a large subset of the Solidity [16] programming language, comparable to the subsets commonly recommended for writing common smart contracts. Lolisa not only includes nearly all the characteristic components of Solidity, such as mapping, modifier, contract, and address types, but it also contains general-purpose programming language features, such as multiple return values, pointer arithmetic, struct, field access and the full power of functions, including recursive functions and function pointers. The main omissions are floating datatype, explicit ether unit of Solidity, the goto statement and non-structured forms of switch such as Duff's device [17]. Particularly, the formal syntax of Lolisa is defined with generalized algebraic datatypes (GADTs) [2] theory, which gives imparts static type annotation to all the values and expressions of Lolisa. In this way, Lolisa has a stronger static typing judgements system than Solidity for checking the construction of programs. As such, it is impossible to construct ill-typed terms in Lolisa. In addition, because the Solidity can equivalently translated into Lolisa directly, the static type-checking system also assists in discovering ill-typed terms in Solidity source code.

The semantics of Lolisa is formally defined in natural semantics. Because Lolisa is employed as the equivalent intermediate language for Solidity, which should be able to be parsed, executed and verified in Coq or similar proof assistants, the semantics of Lolisa are deterministic and base on the GERM framework.

Finally, Lolisa has completely mechanized into Coq. To our knowledge, Lolisa is the first mechanized and validated formal syntax and semantics developed for Solidity.

*4.4. Formal verified interpreter*

The last piece of the puzzle for constructing the FSPVM-E for automatically verifying and symbolically executing simultaneously is the respective formal verified interpreter of Lolisa, denoted as FEther, which is built on the GERM framework and specifications of Lolisa.

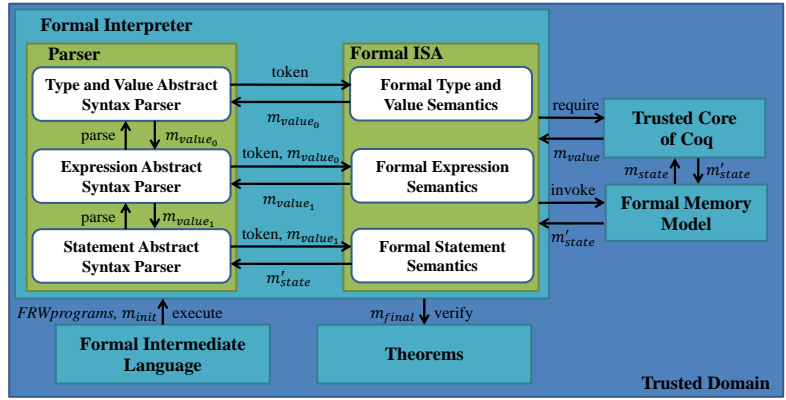

Figure 4. Architecture of the FEther framework

Specifically, as the architecture of FEther shown in Figure 4, the FEther is constructed by two parts: the parser and formal instruction set architecture (ISA). First of all, the parser is employed to analyze the formal abstract syntax of FRWprograms line-by-line separate the tokens of the current executing statement. Second, in essence, the ISA is the formal computational semantics $\mathcal{ES}$ using denotational semantics, which is strictly equivalent with the inductive semantics $\mathcal{S}$ of Lolisa. And the instruction set $\mathcal{Ins}$ of Lolisa is defined as rule 13. The FEther will be adopted to simultaneously execute and derive the FRWprograms state in TCOC. A new intermediate memory state $m_{state}$ will be generated and returned to the formal interpreter as a new $m_{init}$ for the next execution iteration. And the FEther will repeat these process until it is satisfied certain conditions, such as the programs stopped normally or breaking off, and output the final memory state $m_{final}$.

$$\mathcal{Ins} \stackrel{\text{def}}{=} \forall i \in \mathbb{N}. \left( \bigcup_{i=0}^{n} \mathcal{S}_i \right) \leftrightarrow \left( \bigcup_{i=0}^{n} \mathcal{ES}_i \right) \tag{13}$$

Another very important feature of FEther is the combination of gas mechanism of Ethereum virtual machine (EVM) and Bounded Model Checking (BMC) [20]. To be specific, first of all, the combination of symbolic execution and higher-order theorem proving facilitates our use of BMC to verify FRWprograms. We employ BMC notion to set a limitation into the implementation of FEther that it only can execute K times for avoiding the infinite execution situation. Actually, EVM also has the gas mechanism to avoid the same cases in digital transaction. Therefore, the implementation of FEther combines the BMC and gas mechanism. And the execution semantics of FEther follows rule 14 and 15, where the context of the formal memory space is denoted as $M$, $\sigma$ represents the current memory state, the context of the execution environment is represented as $\Omega$ and $\mathcal{F}$ represents the formal system of verification. In addition, the initial environment $env$, which are initialized by the helper function $init_{env}$, and the initial gas value of $env$ is set by $set_{gas}$. These rules represent two conditions of Lolisa programs $P(stt)$ execution, denoted as $\Downarrow_{P(stt)}$. Under the first condition governed by the rule 14, $\Downarrow_{P(stt)}$ terminates after a finite number of steps, which is under K times limitation. Under the second condition governed by the rule 15, $\Downarrow_{P(stt)}$ cannot terminate via its internal logic, and would undergo an infinite number of steps. Therefore, $P(stt)$ is deliberately stopped via the gas-limitation-checking mechanism. Here, $opars$ represents an optional arguments list. The total workload of FEther is about 7000 Coq lines.

$$\frac{\Omega \vdash env \quad M \vdash \sigma \quad \mathcal{F} \vdash opars \quad \Omega, M, \mathcal{F} \vdash P(stt)}{env = set_{gas}\left(init_{env}(P(stt))\right)} \\ \sigma = init_{mem}(P(stt)) \\ \overline{\Omega, M, \mathcal{F} \vdash \langle \sigma, env, opars, \Downarrow_{P(stt)} \rangle \xRightarrow{execute,T} \langle \sigma', env, fenv \rangle} \tag{14}$$

$$\frac{\begin{array}{c} \Omega \vdash env \quad M \vdash \sigma, b_{infor} \quad \mathcal{F} \vdash opars \quad \Omega, M, \mathcal{F} \vdash P(stt) \\ env = set_{gas}\big(init_{env}(P(stt))\big) \quad fenv = init_{env}(P(stt)) \\ \sigma = init_{mem}(P(stt)) \end{array}}{\Omega, M, \mathcal{F} \vdash \langle \sigma, env, opars, \Downarrow_{P(stt)} \rangle \xRightarrow{execute, \infty} \langle \sigma', env' \rangle \land env'.(gas) \leq gasLimit \xRightarrow{T} \langle \sigma', env' \rangle} \tag{15}$$

*4.5. Case study*

To demonstrate the power of our new tool, we have applied our FSPVM-E to specify and verify Smart Sponsor Contract (SSC) [19] again in the Coq proof assistant. As shown in Figure 5, FSPVM-E is already can be used to verify Ethereum smart contracts. In our past work [18], we have verified 6 key theorems of SSC using standard HOLTP approach, and the manual workload is about 1200 Coq lines. Presently, the manual workload of the verification for the same 6 theorems is only 283 Coq lines using FSPVM-E. Obviously, FSPVM-E makes the verification process become much more efficiently. Besides, we also have employed FSPVM-E to verify a number of complex smart contracts, and the relevant examples have been introduced in [5]. In short, the current version of FSPVM-E has already support ERC20 standard, and it is able to almost automatically verify the smart contracts which are developed following the standard.

```
m_0x0000005f := initData;
m_0x00000060 := initData;
m_0x00000061 := initData;
m_0x00000062 := initData;
m_0x00000063 := initData |} (Some nil) (Evn 1 o a0 d) (Evn 1 o a0 d)
(Var (Some public) (Evar (Some close) Tuint);;
 Var (Some public) (Evar (Some quota) Tuint);;
 Var (Some public) (Evar (Some rate) Tuint);;
 Var (Some public) (Evar (Some partiLimit) Tuint);;
 Var (Some public) (Evar (Some totalLimit) Tuint);;
 Var (Some public) (Evar (Some finalLimit) Tuint);;
 If
   (Econst (Vmap priviledges (Mstr_id Iaddress msg (sender ~>> \\\)) None))
   (Assignv (Evar (Some open) Tuint) (Evar (Some privilegeOpen) Tuint);;
    Assignv (Evar (Some close) Tuint) (Evar (Some privilegeClose) Tuint);;
    Assignv (Evar (Some quota) Tuint) (Evar (Some privilegeQuota) Tuint);;
    Assignv (Evar (Some rate) Tuint) RATE_PRIVILEGE;; nil)
   (Assignv (Evar (Some open) Tuint) (Evar (Some ordinaryOpen) Tuint);;
    Assignv (Evar (Some close) Tuint) (Evar (Some ordinaryClose) Tuint);;
    Assignv (Evar (Some quota) Tuint) (Evar (Some ordinaryQuota) Tuint);;
    Assignv (Evar (Some rate) Tuint) RATE_ORDINARY;; nil);;
 If
   (Evar (Some now) Tuint (<) Evar (Some open) Tuint
    (||) Evar (Some now) Tuint (>) Evar (Some close) Tuint) (Throw;; nil)
   (Snil;; nil);;
 If
   (Evar (Some subscription) Tuint (+) Evar (Some rate) Tuint
    (>) TOKEN_TARGET_AMOUNT) (Throw;; nil) (Snil;; nil);;
 If (Evar (Some index) Tuint (==) Econst (Vint (INT I64 Unsigned 0)))
   (Throw;; nil) (Snil;; nil);;
 If
   (Econst (Vmap deposits (Mvar_id Iuint index) None)
```

Figure 5. Formal memory states during the execution and verification of the Lolisa program

## 5. Conclusion and future work

In this paper, we briefly introduce a formal symbolic process virtual machine, denoted as FSPVM-E, which is built on our present research works, for automatically formalizing and verifying Ethereum smart contracts using HOLTP in Coq. Specifically, FSPVM-E is entirely developed in Coq. And FSPVM-E takes execution-verification isomorphism (EVI) as the fundamental theoretical framework, the formal memory framework GERM as the low-level memory model, Lolisa as the source language and FEther as the virtual execution engine to symbolically execute formal version smart contracts in FSPVM-E. Current version of FSPVM-E has already supported the ERC20 standard of Ethereum and can significantly improve the proof efficiency compared with the standard higher-order logic theorem proving approaches.

The FSPVM-E experiment described in this paper is still ongoing, and much work remains to be done: certify the correctness of FEther; optimize the evaluation and proof efficiency of FSPVM-E; extend Lolisa to support more general-purpose programming language; extend the specifications of GERM to shared-memory concurrency; etc. However, the preliminary results obtained so far provide strong evidence that the initial blueprint of building a verified, general and extensible formal symbolic process virtual machine using higher-order logic theorem proving with high-level automation and reusability and without inconsistency problem can be achieved.

**Acknowledgements**

The authors thank Marisa for their kind assistance and LetPub for its linguistic assistance during the preparation of this manuscript.